\documentstyle[12pt,epsf,epsfig]{article}

\def\begeq{\begin{equation}}
\def\endeq{\end{equation}}
\def\begdis{\begin{displaymath}}
\def\enddis{\end{displaymath}}

\begin{document}

\begin{center}
{\bf RAMPED-INDUCED STATES IN THE PARAMETRICALLY DRIVEN GINZBURG-LANDAU MODEL}

Boris A. Malomed

Department of Interdisciplinary Studies, Faculty of Engineering, Tel Aviv
University, Tel Aviv 69978, Israel

Horacio G. Rotstein

Department of Chemistry and Volen Center for Complex Systems, Brandeis
University, MS 015 Waltham, MA 02454-9110, USA

\bigskip

{\bf Abstract}
\end{center}

We introduce a parametrically driven Ginzburg-Landau (GL) model, which
admits a gradient representation, and is subcritical in the absence of the
parametric drive (PD). In the case when PD acts uniformly in space, this
model has a stable kink solution. A nontrivial situation takes places when
PD is itself subject to a kink-like spatial modulation, so that it selects
real and imaginary constant solutions at $x=\pm \infty $. In this situation,
we find stationary solutions numerically, and also analytically for a
particular case. They seem to be of two different types, viz., a pair of
kinks in the real and imaginary components, or the same with an extra kink
inserted into each component, but we show that both belong to a single
continuous family of solutions. The family is parametrized by the coordinate
of a point at which the extra kinks are inserted. However, solutions with
more than one kink inserted into each component do not exist. Simulations
show that the former solution is always stable, and the latter one is, in a
certain sense, neutrally stable, as there is a special type of small
perturbations that remain virtually constant in time, rather than decaying
or growing (they eventually decay, but extremely slowly).

\newpage

\section{Introduction}

Various forms of parametrically driven Ginzburg-Landau (GL) equations
constitute a class of pattern-forming systems with broken intrinsic
phase-rotation invariance \cite{Coullet,Barash}. In this work, we consider a 
{\it gradient} model of this type, 
\begin{equation}
u_{t}=\epsilon (x)\cdot u^{\ast }+\zeta u+u_{xx}-|u|^{2}u\equiv -\frac{%
\delta L}{\delta u^{\ast }},  \label{1}
\end{equation}
where $u^{\ast }$ stands for the complex conjugation, $\delta /\delta
u^{\ast }$is the variational derivative, the function $\epsilon (x)$ and
constant $\zeta $ are real, and the {\it Lyapunov functional} is 
\begin{equation}
L=\frac{1}{2}\int_{-\infty }^{+\infty }[-\epsilon (x)\cdot \left(
u^{2}+(u^{\ast })^{2}\right) -\zeta |u|^{2}+|u_{x}|^{2}+|u|^{4}]dx.
\label{L}
\end{equation}
Equation (1) with $\epsilon ={\rm const}$ has various applications, the
best-known one being the Rayleigh-B\'{e}nard convection in the case when the
convection-driving temperature difference across the fluid layer is subject
to a resonant periodic modulation along the layer (the coordinate $x$ is
running along the layer), or the boundary is similarly undulated, see early 
\cite{Alik} and more recent \cite{Zimmer} theoretical works on this topic,
and an experimental work \cite{Busse}. The ``resonant'' character of the
modulation means that its period is half the depth of the convective layer.
The presence of the variable coefficient $\epsilon (x)$ implies that the
``primary'' modulation is further subject to a ``supermodulation'' on a
scale essentially larger than the layer's depth, which is quite possible in
large-aspect-ratio convection cells, and gives rise to novel patterns, as it
will be shown below.

The real constant $\zeta $ in Eq. (1) controls the system's overcriticality
in the absence of the parametric drive: when $\epsilon =0$, the situation is 
{\it over}critical if $\zeta >0$, and {\it sub}critical if $\zeta <0$.
Rescaling allows one to reduce $\zeta $ to one of the three values, $\zeta
=-1,0,\,$or${\rm \,}+1$. To the best of our knowledge, the previous works\
were dealing with the case when the gradient GL system was overcritical when 
$\epsilon =0$. Our aim is to consider the opposite case with $\zeta =-1$,
when no nontrivial pattern may exist without the parametric drive. Thus, $%
\zeta \equiv -1$ will be assumed below, unless another value of $\zeta $ is
specified.

In the case $\epsilon ={\rm const}$, Eq. (1) has obvious constant solutions, 
\begin{equation}
u_{0}=\pm \sqrt{\epsilon -1},\,{\rm or}\,\,u_{0}=\pm i\sqrt{-\epsilon -1},
\label{constant}
\end{equation}
which exist provided that, respectively, $\epsilon >1$ or $\epsilon <-1$,
while the trivial solution $u=$ $0$ is stable in the case $|\epsilon |\leq 1$
, when the nonzero solutions (\ref{constant}) do not exist. In the case $%
|\epsilon |>1$, another available exact solution is a {\it kink}, or domain
wall, which provides for a transition between the constant solutions (\ref
{constant}) with the opposite signs. For instance, in the case $\epsilon >1$
(recall we set $\zeta =-1$), the kink solution to Eq. (1) is 
\begin{equation}
u_{{\rm kink}}(x)=\sigma \sqrt{\epsilon -1}\tanh \left( x\sqrt{\frac{%
\epsilon -1}{2}\,}\right) \,\,,  \label{Neel}
\end{equation}
where $\sigma =\pm 1$ is the kink's polarity.

Note that a similar kink, with $(\epsilon -1)$ replaced by $1$ is,
simultaneously, an exact solution to the usual GL equation, which can be
obtained from Eq. (1) setting $\epsilon =0$ and $\zeta =+1$. It is commonly
known that, in its latter capacity, the real kink solution is{\em \ }%
unstable against purely imaginary perturbations. On the contrary to this, an
elementary consideration demonstrates that the solution (\ref{Neel}) is
stable as a solution to the parametrically driven GL equation (1) with $%
\zeta =-1$ and $\epsilon >1$. This is a formally new result, but the actual
objective of the work is to consider more interesting solutions in the case
when the parametric drive is {\it ramped}, i.e., it is itself subject to a
kink-like modulation, see below.

\section{The ramp and stationary solutions}

A configuration with $\epsilon (x)$ monotonically varying between
asymptotically constant values at $x=\pm \infty $ is called a {\it ramp}. In
the case of the usual (not parametrically driven) GL equation, much work has
been done for a case when the overcriticality is ramped so that the system
is subcritical at $x=-\infty $ and supercritical at $x=+\infty $, as this
configuration makes it possible to uniquely solve the {\it %
wavenumber-selection problem}, and demonstrates other remarkable features 
\cite{ramp}. Moreover, the ramp problem can be extended into the
two-dimensional geometry, replacing $u_{xx}$ by the two-dimensional
Laplacian. In particular, the stability of the ramp-supported
quasi-one-dimensional pattern in the two-dimensional version of the usual GL
equations was investigated in Refs. \cite{we}. In the parametrically driven
equation (1) with $\zeta =-1$, a similar ramp can easily be arranged too,
choosing a function $\epsilon (x)$ which is monotonically varying between a
value $|\epsilon (-\infty )|<1$, that provides for the stability of the $u=0$
solution at $x=-\infty $, and $|\epsilon (+\infty )|>1$, which gives rise to
a stable nonzero solution (\ref{constant}) at $x=+\infty $. 

The subject of this work is to consider a more nontrivial case of the ramp,
when $\epsilon (x)$ interpolates between values $\epsilon (-\infty )<-1$ and 
$\epsilon (\infty )>+1$, for instance 
\begin{equation}
\epsilon (x)=a\,\tanh \left( \lambda x\right) \,,\,\,a>1,\,\,\lambda >0.
\label{odd}
\end{equation}
This ramp supports purely imaginary and purely real nonzero states (\ref
{constant}) at $x=-\infty $ and $x=+\infty $, so that $u(x=-\infty )=\pm i%
\sqrt{a-1}\,$, $u(x=+\infty )=\pm \sqrt{a-1}$. Our objective is to find
solutions interpolating between these asymptotic states, and investigate
their stability. A configuration of the present type can be easily
implemented experimentally, which makes the problem physically relevant (in
the convective layer, the constant imaginary and real solutions represent
uniform arrays of convection rolls mutually shifted by half the layer's
depth).

For the numerical solution of Eq. (1) with $\zeta =-1$ and $\epsilon (x)$
taken as per Eq. (\ref{odd}), we define $u(x)\equiv v(x)+iw(x)$, arriving at
a system of equations for two real functions, 
\begin{eqnarray}
v_{t} &=&a\,\tanh \left( \lambda x\right) \cdot v+v_{xx}-\left(
v^{2}+w^{2}+1\right) v=0\,,  \label{v} \\
w_{t} &=&-a\,\tanh \left( \lambda x\right) \cdot w+w_{xx}-\left(
v^{2}+w^{2}+1\right) w=0\,.  \label{w}
\end{eqnarray}
Boundary conditions (b.c.) corresponding to the situation defined above are 
\begin{eqnarray}
v(x=-\infty )=0,\ \ \ \ \ \ \ \ \ \ \ &&\,\,v(x=+\infty )=\pm \sqrt{a-1},
\label{boundary_v} \\
w(x=-\infty )=\pm \sqrt{a-1}\,, &&\,w(x=+\infty )=0,\ \ \ \ \ \ \ \ \ \ 
\label{boundary_w}
\end{eqnarray}
the signs $\pm $ in Eqs. (\ref{boundary_v}) and (\ref{boundary_w}) being
mutually independent. Obvious symmetries of the equations allow us to fix
only one set of the signs, as solutions corresponding to other combinations
of the signs are actually the same.

Getting back for a moment to Eq. (1) with $\zeta =+1$, it is worthy to note
that, in this case, a particular choice $a=1$ and $\lambda =1/\sqrt{2}$
gives rise to an {\em exact} stationary solution interpolating between the
real and imaginary asymptotic states, viz., 
\begin{equation}
v(x)=\left( 1/\sqrt{2}\right) \left[ 1+\tanh \left( x/\sqrt{2}\right) \right]
,\,\,\,w(x)=\left( 1/\sqrt{2}\right) \left[ 1-\tanh \left( x/\sqrt{2}\right) %
\right] .  \label{particular}
\end{equation}
However, if $\zeta =-1$, no similar exact solution can be found. There is
only one analytically tractable (and rather unphysical) case, which is $%
\lambda \ll 1$, i.e., a very broad ramp. In this case, defining $\xi \equiv
\lambda x$, one can readily see that the $x$-derivative terms in Eqs. (\ref
{v}) and (\ref{w}) are negligible, and an asymptotically exact stationary
solution corresponding to $\lambda \rightarrow 0$ is 
\begin{eqnarray}
w &=&0,\,v^{2}=a\tanh \xi -1\ \ {\rm at}\ \ \tanh \xi >1/a,  \nonumber \\
v &=&0,\,w^{2}=-a\tanh \xi -1\ \ {\rm at}\ \ \tanh \xi <-1/a,  \nonumber \\
v &=&w=0\,\,{\rm at}\,\,\left| \tanh \xi \right| <1/a.  \label{primitive}
\end{eqnarray}
Of course, at small but finite $\lambda $ the fields which exactly vanish in
various regions in the limit $\lambda =0$ have small nonzero values at all $%
\xi $.

In order to understand the structure of solutions to Eqs. (\ref{v}) and (\ref
{w}), we consider the following auxiliary system, 
\begin{equation}
\left\{ 
\begin{array}{l}
v_{xx}-(\alpha ^{2}\ v^{2}+\alpha ^{2}\ w^{2}+1)\ v+a\ g(x)v\ =0, \\ 
\\ 
w_{xx}-(\alpha ^{2}\ v^{2}+\alpha ^{2}\ w^{2}+1)\ w-a\ g(x)w\ =0,
\end{array}
\right.  \label{eq:ecuaciones}
\end{equation}
subject to b.c.

\begin{equation}
\left\{ 
\begin{array}{lll}
v(-\infty )=0, &  & v(\infty )=1, \\ 
w(-\infty )=1, &  & w(\infty )=0.
\end{array}
\right.   \label{eq:borde}
\end{equation}
Here $\alpha =\sqrt{a-1}$, and 
\begin{equation}
g(x)\equiv \left\{ 
\begin{array}{lll}
1, &  & x\in \lbrack \eta ,\infty ) \\ 
x/\eta , &  & x\in \lbrack -\eta ,\eta ] \\ 
-1, &  & x\in (-\infty ,-\eta ]
\end{array}
\right. ,  \label{eq:defg}
\end{equation}
with $0<\eta \ll 1$. The system (\ref{eq:ecuaciones}) reduces to Eqs. (\ref
{v})-(\ref{w}) if one substitutes $g(x)$ by $\tanh (\lambda x)$,
simultaneously replacing $v$ and $w$ by $v/\alpha $ and $w/\alpha $ in Eqs. (%
\ref{v})-(\ref{w}).

It is easy to see that, due to symmetry arguments, $v(x)=w(-x)$ and $%
v(-x)=w(x)$, which in particular means that $v(0)=w(0)$. Far from the
origin, obvious {\it outer-zone} solutions to Eqs. (\ref{eq:ecuaciones})
satisfying b.c. (\ref{eq:borde}) are 
\begin{equation}
\left\{ 
\begin{array}{llll}
v=\pm 1, & w=0, & {\rm at} & x\geq \eta , \\ 
v=0, & w=\pm 1, & {\rm at} & x\leq -\eta .
\end{array}
\right.   \label{eq:sol0}
\end{equation}
Inside the interval $x\in \lbrack -\eta ,\eta ]$ one has, to the leading
order in the small parameter $\eta $, $v_{xx}=w_{xx}=0$. An {\it inner-zone}
solution to these equations, satisfying b.c. $v(\eta )=1$, $v(-\eta )=0$, $%
w(\eta )=0$ and $w(-\eta )=1$, which are necessary for the continuous
matching to the outer-zone solution (\ref{eq:sol0})with the upper sign, is

\begin{equation}
v=\frac{x}{2\ \eta }+\frac{1}{2}\,,\ \ w=-\frac{x}{2\ \eta }+\frac{1}{2}\ .\
\ \ \ \ \ \ \ \   \label{eq:sol1}
\end{equation}

\begin{figure}[ph]
\begin{tabular}{ll}
\epsfig{file=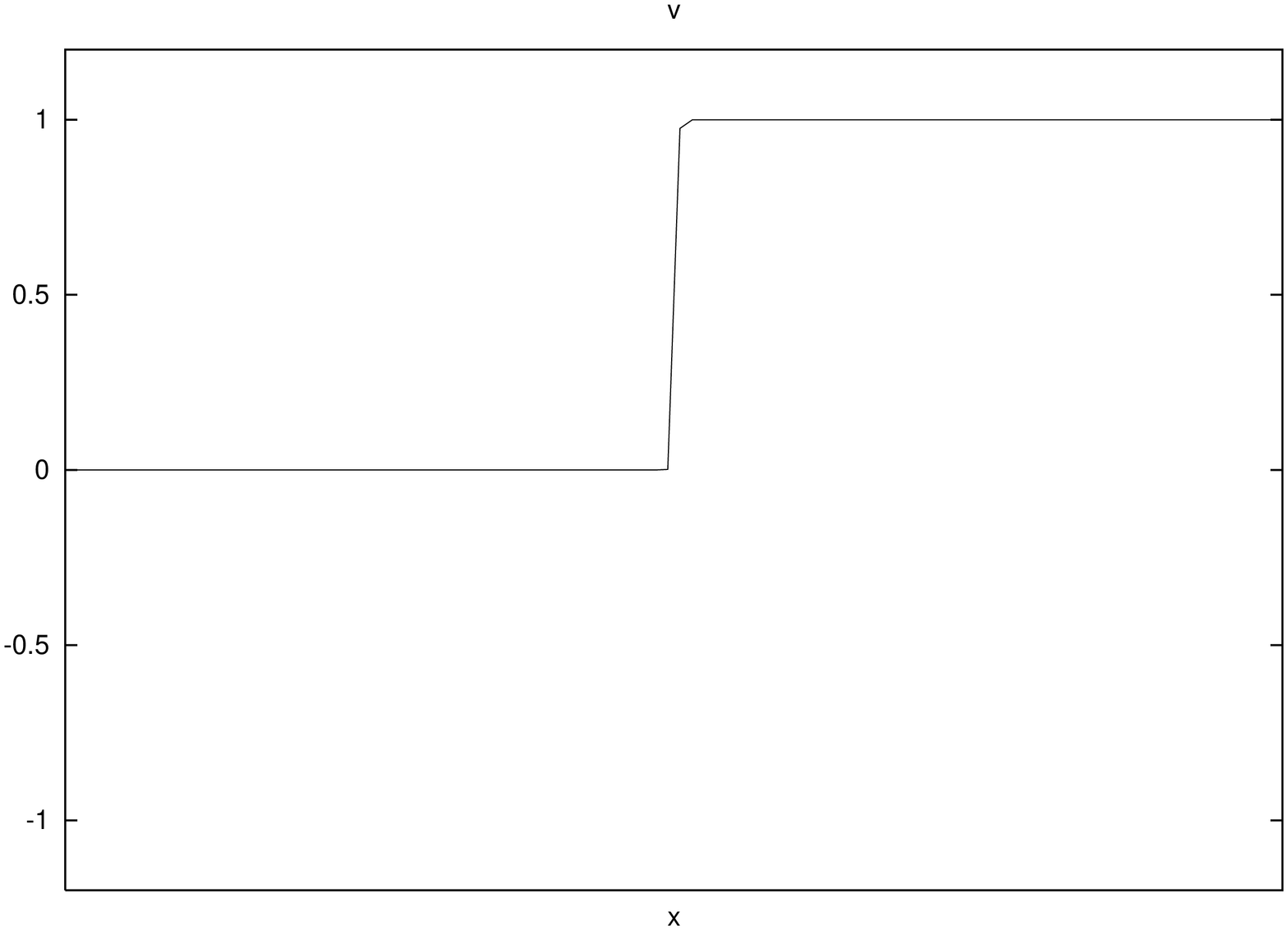,height=12cm,width=6cm,angle=-90} &  \\ 
\epsfig{file=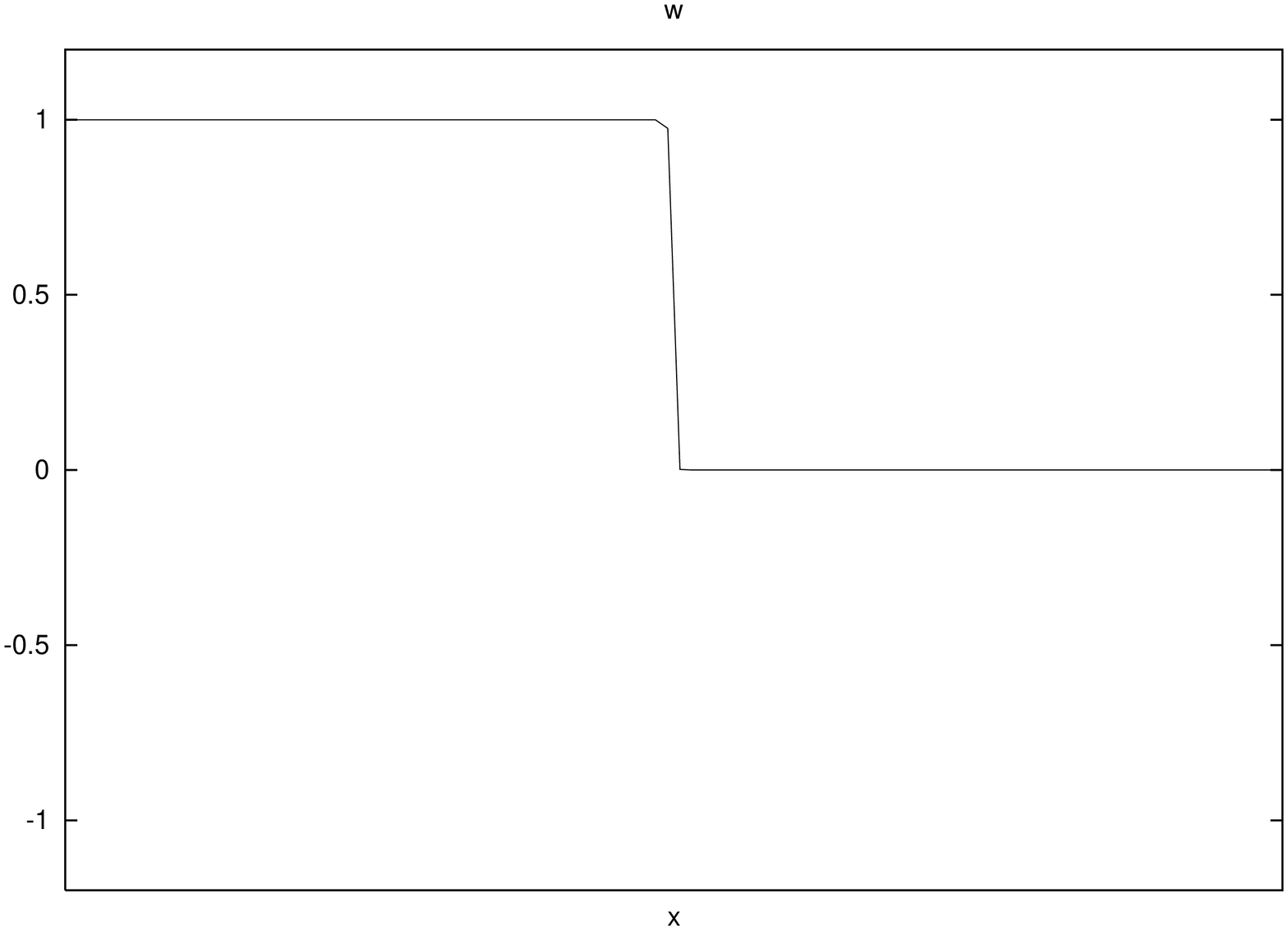,height=12cm,width=6cm,angle=-90} & 
\end{tabular}
\caption{A general scheme of the type-A solution. The corners of the kinks in the
$v$ and $w$ components are rounded since the exact solution, unlike the approximation
(\ref{eq:sol2}), does not admit a discontinuity of the first derivatives.}
\label{sol_A}
\end{figure}

Thus, to the leading order, a solution to the auxiliary equations (\ref
{eq:ecuaciones})-(\ref{eq:borde}), that we will call a {\it type-A}
solution, is

\begin{equation}
\begin{array}{l}
v(x)=\left\{ 
\begin{array}{lll}
0, &  & x\in (-\infty ,-\eta ], \\ 
(1+x/\eta )/2, &  & x\in \lbrack -\eta ,\eta ] \\ 
1, &  & x\in \lbrack \eta ,\infty ),
\end{array}
\right.  \\ 
\\ 
w(x)=\left\{ 
\begin{array}{lll}
(1-x/\eta )/2, &  & x\in \lbrack -\eta ,\eta ], \\ 
1, &  & x\in (-\infty ,-\eta ], \\ 
0, &  & x\in \lbrack \eta ,\infty ).
\end{array}
\right. 
\end{array}
\label{eq:sol2}
\end{equation}
An example of the type-A solution is displayed in Fig. (\ref{sol_A}).

We note that there also exists another inner-zone solution to the above
leading-order equations $v_{xx}=w_{xx}=0$, that satisfies b.c. $v(\eta )=-1$%
, $v(-\eta )=0$, $w(\eta )=0$ and $w(-\eta )=-1$:

\begin{equation}
v=-\frac{x}{2\ \eta }-\frac{1}{2}\ ,\ w=\frac{x}{2\ \eta }-\frac{1}{2}\ .\ \
\ \ \ \ \ \ \   \label{eq:sol3}
\end{equation}
This inner solution is matched to the outer one (\ref{eq:sol0}) with the
lower sign. In order to make the latter solution also consistent with b.c. (%
\ref{eq:borde}), we need to connect the two constant solutions $v=\pm 1$ in
the outer region $x>\eta $. To this end, we consider a region at $x>\eta $
where $w=0$ and $g(x)=1$. Then the second equation in the system (\ref
{eq:ecuaciones}) is automatically satisfied, and the first equation yields 
\begin{equation}
v_{xx}-(a-1)\ (v^{3}-v)=0.  \label{eq:ecua2}
\end{equation}

\begin{figure}[ph]
\begin{tabular}{ll}
\epsfig{file=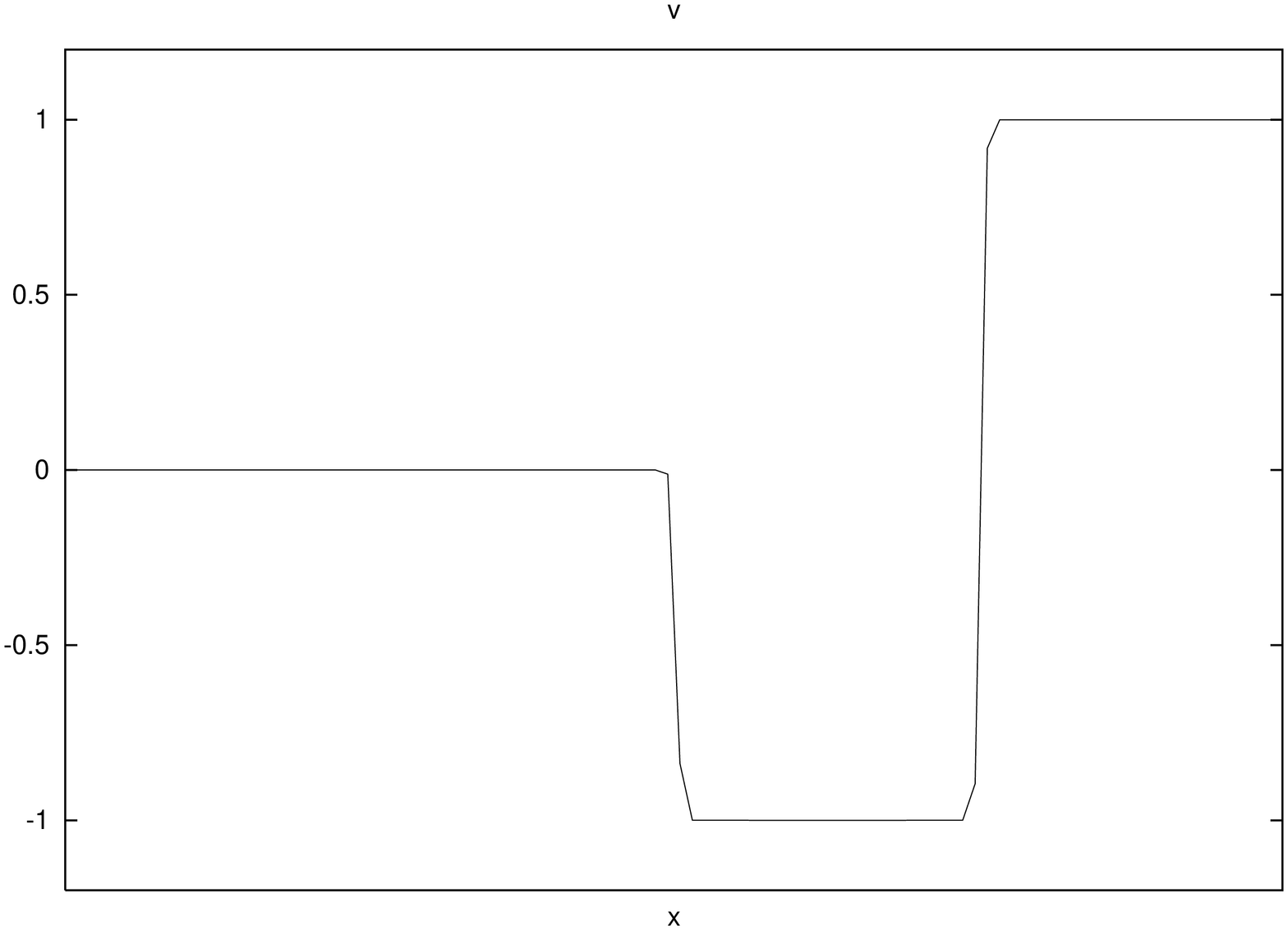,height=12cm,width=6cm,angle=-90} &  \\ 
\epsfig{file=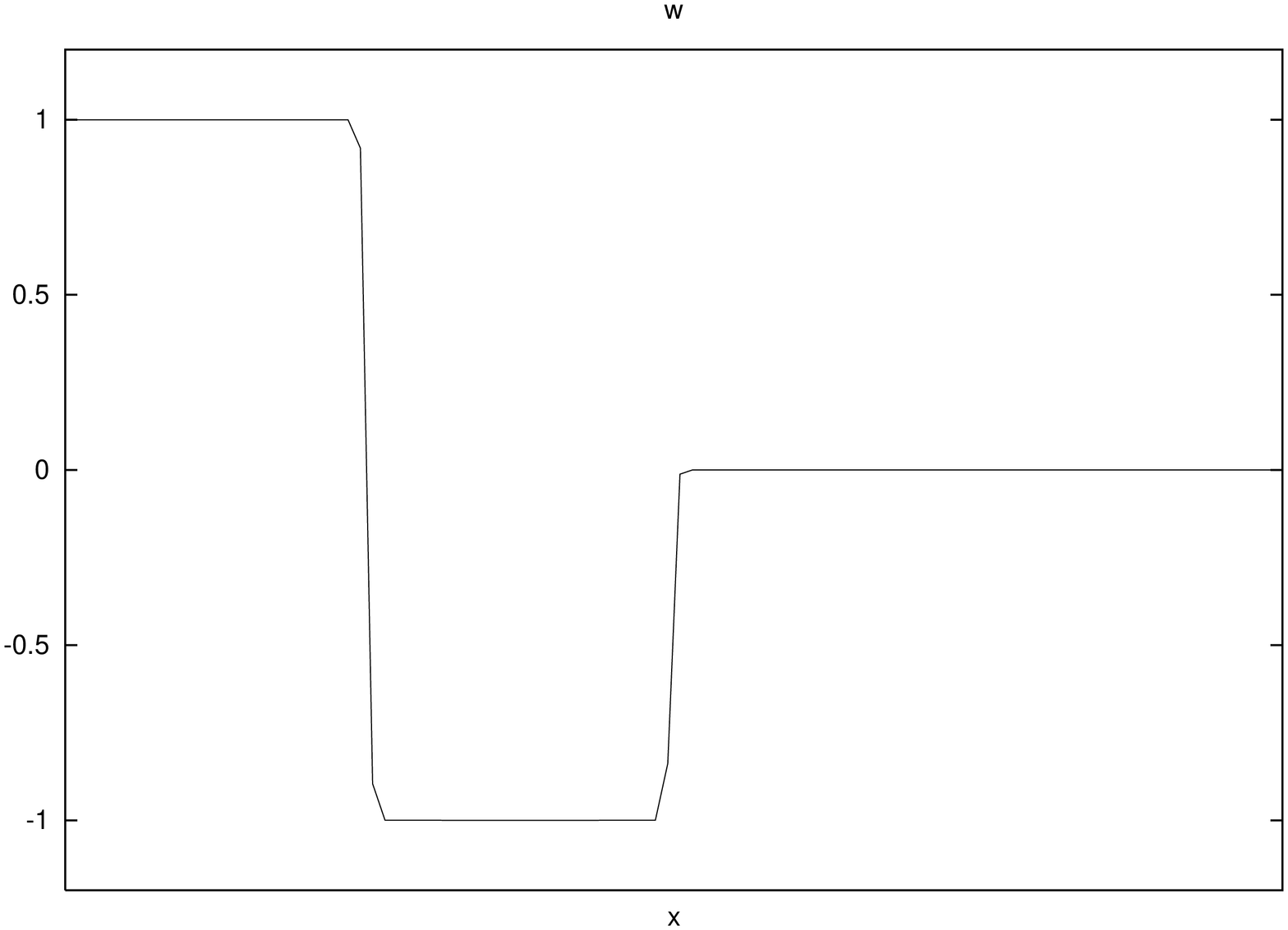,height=12cm,width=6cm,angle=-90} & 
\end{tabular}
\caption{A general scheme of the type-B solution.}
\label{sol_B}
\end{figure}

We choose an arbitrary point $x_{0}>2\eta $ and look for a solution to Eq. (%
\ref{eq:ecua2}) in an interval $(x_{0}-\eta ,x_{0}+\eta )$. If $(a-1)\ \eta
^{2}\ll 1$, then the solution is $v=(x-x_{0})/\eta $. If $(a-1)\ \eta ^{2}=%
{\cal O}(1)$, then the solution to (\ref{eq:ecua2}) is $v=\tanh (\sqrt{%
(a-1)/2}\ (x-x_{0}))$, whose expansion at small $|x-x_{0}|$ is again $%
(x-x_{0})/\eta $. Thus, to the leading order, an approximate solution to
Eqs. (\ref{eq:ecuaciones})-(\ref{eq:borde}) that we call a {\it type-B}
solution, and which is valid for an arbitrary $x_{0}>2\eta $, is

\begin{equation}
\begin{array}{l}
v(x)=\left\{ 
\begin{array}{lll}
0, &  & x\in (-\infty ,-\eta ], \\ 
-(1+x/\eta )/2 &  & x\in \lbrack -\eta ,\eta ], \\ 
-1, &  & x\in \lbrack \eta ,x_{0}-\eta ], \\ 
(x-x_{0})/\eta , &  & x\in \lbrack x_{0}-\eta ,x_{0}+\eta ], \\ 
1 &  & x\in ([x_{0}+\eta ,\infty ).
\end{array}
\right.  \\ 
\\ 
w(x)=\left\{ 
\begin{array}{lll}
1, &  & x\in (-\infty ,-x_{0}-\eta ], \\ 
-(x+x_{0})/\eta , &  & x\in \lbrack -x_{0}-\eta ,-x_{0}+\eta ], \\ 
-1, &  & x\in \lbrack -xo+\eta ,-\eta ] \\ 
(1-x/\eta )/2, &  & x\in \lbrack -\eta ,\eta ], \\ 
0, &  & x\in \lbrack \eta ,\infty ).
\end{array}
\right. 
\end{array}
\label{eq:sol4}
\end{equation}
An example of this solution is displayed in Fig \ref{sol_B}.

It is easy to see that the type-B solution (\ref{eq:sol4}) can also be
represented in the form 
\begin{equation}
\begin{array}{lll}
v(x)=2\ f(-x-x_{0})-f(-x), &  & w(x)=2\ f(x-x_{0})-f(x),
\end{array}
\label{eq:sol5}
\end{equation}
where $f(x)$ is the $v$-component of the type-A solution, i.e.,

\begin{equation}
f(x)\equiv \left\{ 
\begin{array}{lll}
0, &  & x\in (-\infty ,-\eta ], \\ 
(1+x/\eta )/2, &  & x\in \lbrack -\eta ,\eta ] \\ 
1, &  & x\in \lbrack \eta ,\infty ),
\end{array}
\right.   \label{eq:sol6}
\end{equation}
The representation (\ref{eq:sol5}) suggests that there exists a continuous 
{\it family} of solutions to the auxiliary system (\ref{eq:ecuaciones}),
parametrized by $x_{0}$, so that when $x_{0}=0$, the solution is of the type
A, and when $x_{0}>2\epsilon $, the solution is of the type B. In the
intermediate case $x_{0}\in (0,2\eta )$, the solution cannot be represented
in the simple approximate form (\ref{eq:sol5}), and can only be constructed
numerically. In other words, the extra kinks in the $v$- and $w$-
components, which distinguish the solutions of the type B from those of the
type A, can be inserted, essentially, at any point, i.e., there is no unique
``equilibrium'' point at which the extra kinks have to be placed in order to
provide for a stationary solution.

Once the general structure of the solutions to the auxiliary system (\ref
{eq:ecuaciones}) is understood, it is natural to check whether stationary
solutions of the underlying equations (8) and (9) follow the same pattern.
First, in order to check the existence of type-A stationary solutions to
Eqs. (8) and (9), we have performed numerical integration of the
boundary-value problems \cite{numerical} by means of the finite-difference
method in a domain of the size $\Delta x=100$ (as it will be seen from
results displayed below, this size is quite sufficient to consider the
system as being practically infinitely large). The number of spatial points
was $8192$, so that the spatial stepsize was $0.0122$.

The numerical problem was thus reduced to a system of algebraic equations of
the order $2\cdot \left( 8192-1\right) $, which were solved by means of the
Newton's method with tolerance $10^{-4}$. As the first step, we looked for
type-A solutions, taking, as initial trials for $v(x)$ and $w(x)$, functions 
$C/(C+e^{-\gamma \ x})$ and $C/(C+e^{\gamma \ x})$ with different values of $%
C$ and $\gamma $, which seem close to the type-A solutions sought for. The
calculations have been made over a large range of values of the parameters
from Eq. (\ref{odd}), $a=2\,-\,1000$ and $\lambda =0.1\,-\,10000$. In all
cases, a type-A solution has been found indeed.

The next step was to check the existence of the type-B solutions. To this
end, we employed two different tests. First, we computed type-B stationary
solutions by means of the algorithm described above but with initial trial
functions given by the expression (\ref{eq:sol5}), where $f(x)$ was replaced
by the previously found type-A solution, for the same values of $a$ and $%
\lambda $, and various values of $x_{0}$.

Following this way, in most cases tested we could indeed obtain type-B
solutions corresponding to, virtually, arbitrary values of $x_{0}$, in
accord with the results of the analysis performed above. However, the result
was inconclusive, due to numerical problems, for the most intriguing cases
in which $x_{0}$ is such that the minima of the initial trial functions (\ref
{eq:sol5}), with $f(x)$ replaced by the previously found type-A solution,
were not close enough to $-1$ (those values of $x_{0}$ are close to $0$). In
those cases the sequence of approximations produced by the Newton's method
did not converge to a solution sought for within the required tolerance in a
reasonable amount of steps.

The second method, which, simultaneously, gave a direct information about
the dynamical stability of the stationary solutions, was based on simulating
the full system of PDEs (8) and (9), this time using the expression (\ref
{eq:sol5}), with $f(x)$ replaced by the corresponding type-A solution found
at the previous step, as an i{initial configuration}. In doing this, we
again took different values of the parameter $x_{0}$ in Eq. (\ref
{eq:sol5}).

The numerical simulations of the full nonlinear PDEs were performed using a
spectral method with the FFT transform based on $8192$ spatial points, and
the modified Euler's method for the time evolution with the step $\delta
t=3\cdot 10^{-5}$, in a time interval which was checked to be sufficient to
obtain a stationary solution (the largest interval was $0<t<30$).

The values of $x_{0}$ were chosen, in all the cases, as corresponding to $%
400k$ spatial discretization points, $k=1,\ldots ,6$ (i.e., $%
x_{0}=4.883,\,9.766,\,14.648,\,19.531,\,24.414,\,29.297$). We have found
that the solution, essentially, remains almost constant in time. In
particular, the value of $x$ at which $v$ and $w$ cross zero, while varying
from $v=-1$ to $v=1$ and $w=1$ to $w=-1$ respectively, remains virtually
constant. This numerical observations support the conjecture that the
expression (\ref{eq:sol5}) with $f(x)$ replaced by the corresponding type-A
stationary solution is a fairly good approximation to the type-B solution of
Eqs. (\ref{v}) and (\ref{w}).

\begin{figure}[ph]
\begin{tabular}{ll}
\epsfig{file=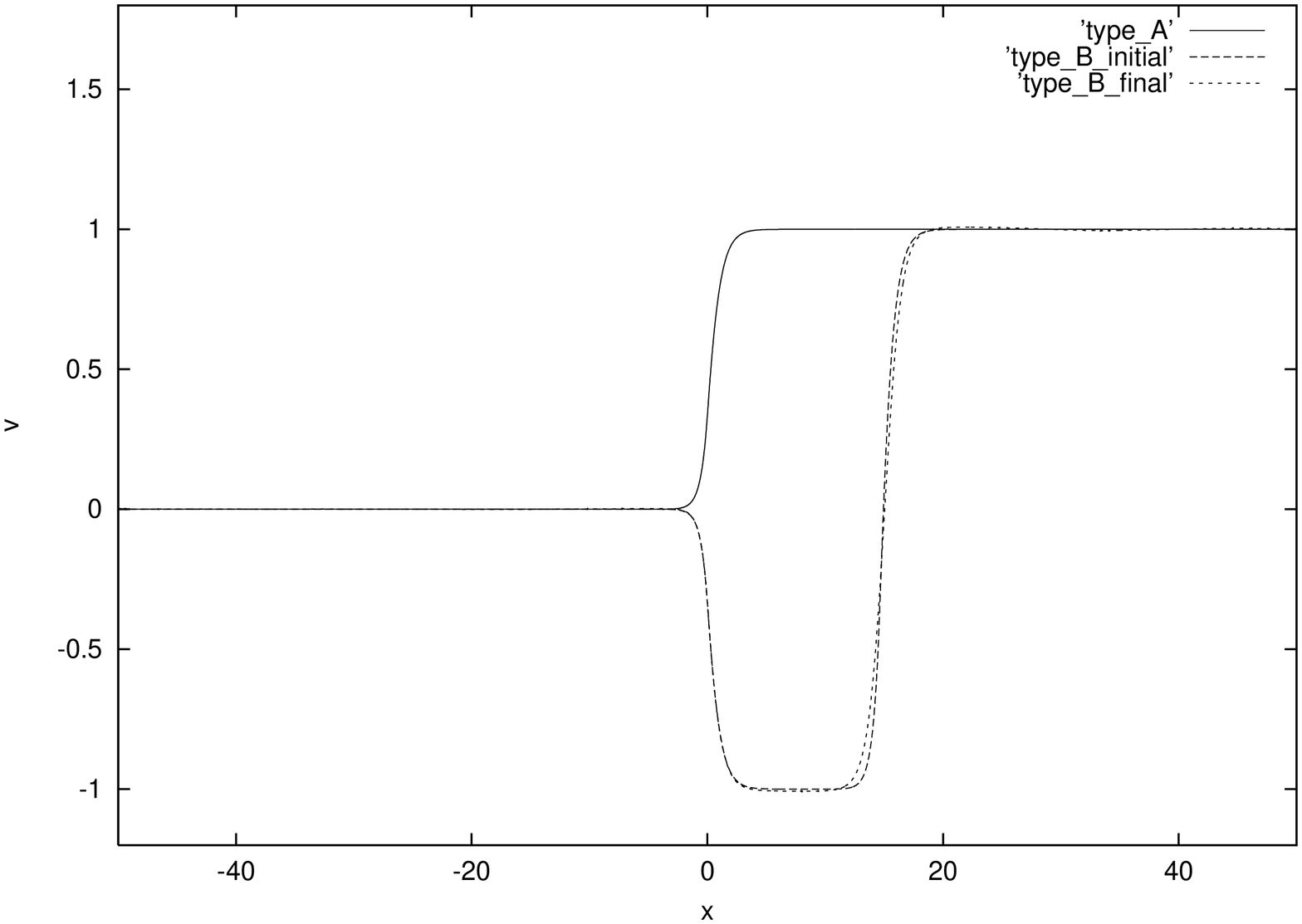,height=12cm,width=6cm,angle=-90} &  \\ 
\epsfig{file=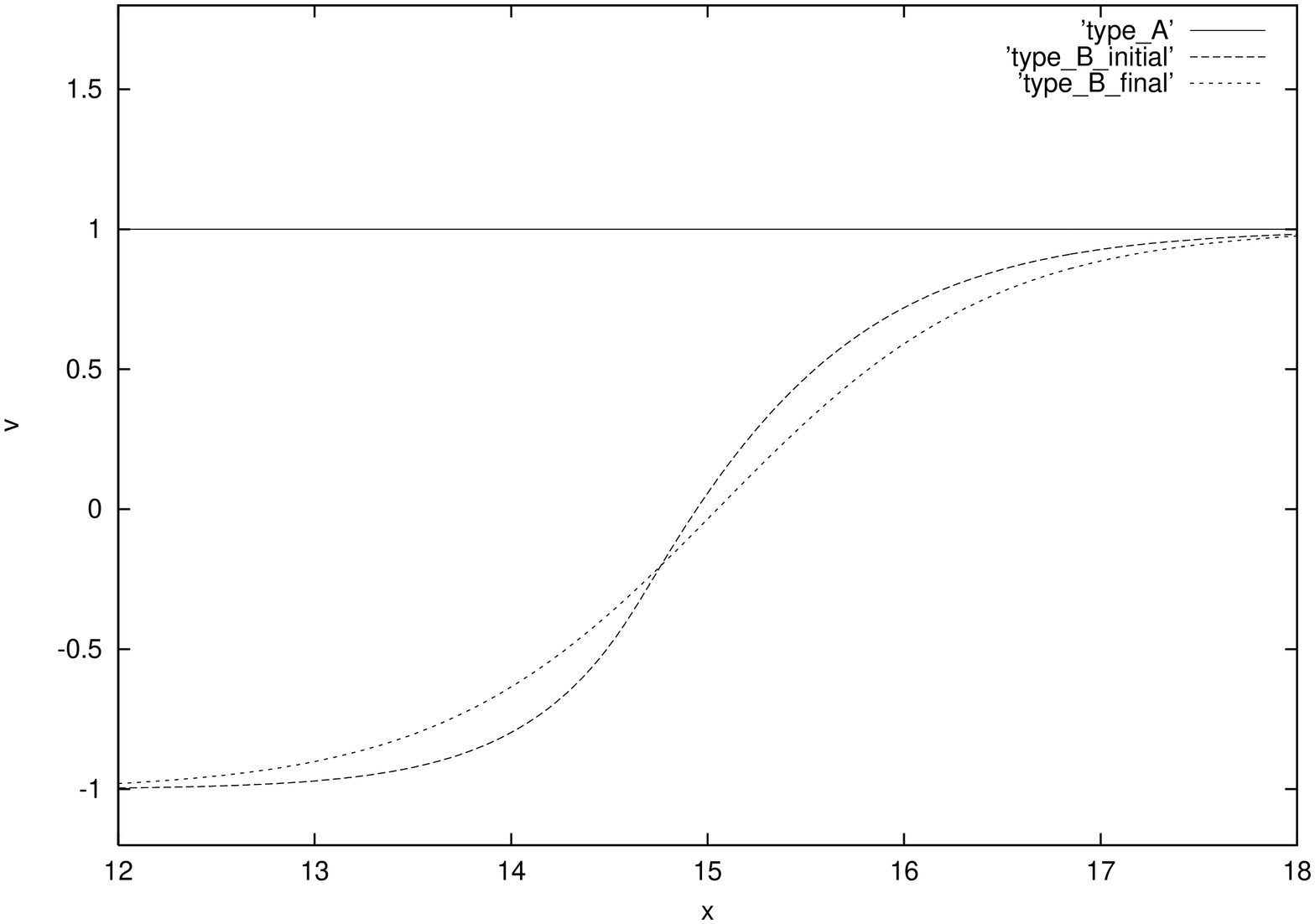,height=12cm,width=6cm,angle=-90} & 
\end{tabular}
\caption{ The $v$- components of the type-A and type-B solutions to Eqs.
(8) and (9). The latter solution was obtained using the initial condition
in the form (\ref{eq:sol5}) with $f(x)$ replaced by the type-A solution,
that was found directly from the stationary version of Eqs. (8) and (9). The
upper panel displays the solution as a whole, while the lower panel is a
blowup of the region where the B-type solution makes its jump. In this
figure, $a = 2$, $\protect\lambda = 10$, and the time moment shown
is $t = 30 $.}
\label{mapad-long-01}
\end{figure}

This approximation has been found to further improve as both $a$ and $%
\lambda $ in Eq. (\ref{odd}) increase. As an illustration, we show in Fig. 
\ref{mapad-long-01} the graphs of the $v$- components of the Type-A and
type-B solutions found for $a=2$ and $\lambda =10$ at the time moment $t=30$%
. The B-type solution was obtained, in accord with what was said above, from
the initial condition in the form of Eq. (\ref{eq:sol5}) with $f(x)$
replaced by the type-A solution that was found directly from the stationary
ODEs.

As well as in the case of the auxiliary system (\ref{eq:ecuaciones}) and (%
\ref{eq:defg}), the existence of the B-type solutions can be qualitatively
understood. Indeed, it is clearly seen in Figs. \ref{sol_A} and \ref{sol_B}
that the abrupt jump between positive and negative values of, say, $v$ takes
place when the $w$-component is practically equal to zero and, moreover, we
may replace $\tanh \left( \lambda x\right) $ by $+1$ or $-1$, as the
above-mentioned jump takes place far from the place where $\tanh \left(
\lambda x\right) $ is essentially different from $\pm 1$. Thus, as a matter
of fact, we are dealing with a very simple stationary GL equation, (\ref
{eq:ecua2}), about which it is well known that it has a single stable
nontrivial solution in the form of a kink, $v=$ $\pm \sqrt{a-1}$ $\tanh
\left( \sqrt{(a-1)/2}x\right) $, cf. Eq. (4). Therefore, the jump in the
type-B solution exactly corresponds to this kink solution. Moreover, it is
known too that more complicated stationary solutions to the real GL
equation, containing several kinks and/or antikinks, do not exist, which
explains why we have never obtained solutions with a larger number of jumps.

\section{Stability of the stationary solutions}

Results presented in the previous section suggest that both the type-A and
type-B solutions are dynamically stable as stationary solutions to the PDE
system. To test the stability in a more systematic way, we simulated the
non-stationary equations (\ref{v}) and (\ref{w}), adding to the (previously
found) stationary solutions small initial perturbations in the form of an
arbitrary combination of spatial harmonics compatible with the boundary
conditions.

Actually, we, first of all, solved the non-stationary equations (\ref{v})
and (\ref{w}) which were l{inearized} around the stationary solutions found
in the previous section. In this approximation, all the small perturbations
added to an arbitrary type-A stationary solution quickly decay. Thus, we
conclude that all these solutions are unequivocally stable.

\begin{figure}[ph]
\begin{tabular}{llllllll}
{\large {\bf (a) }} & %
\epsfig{file=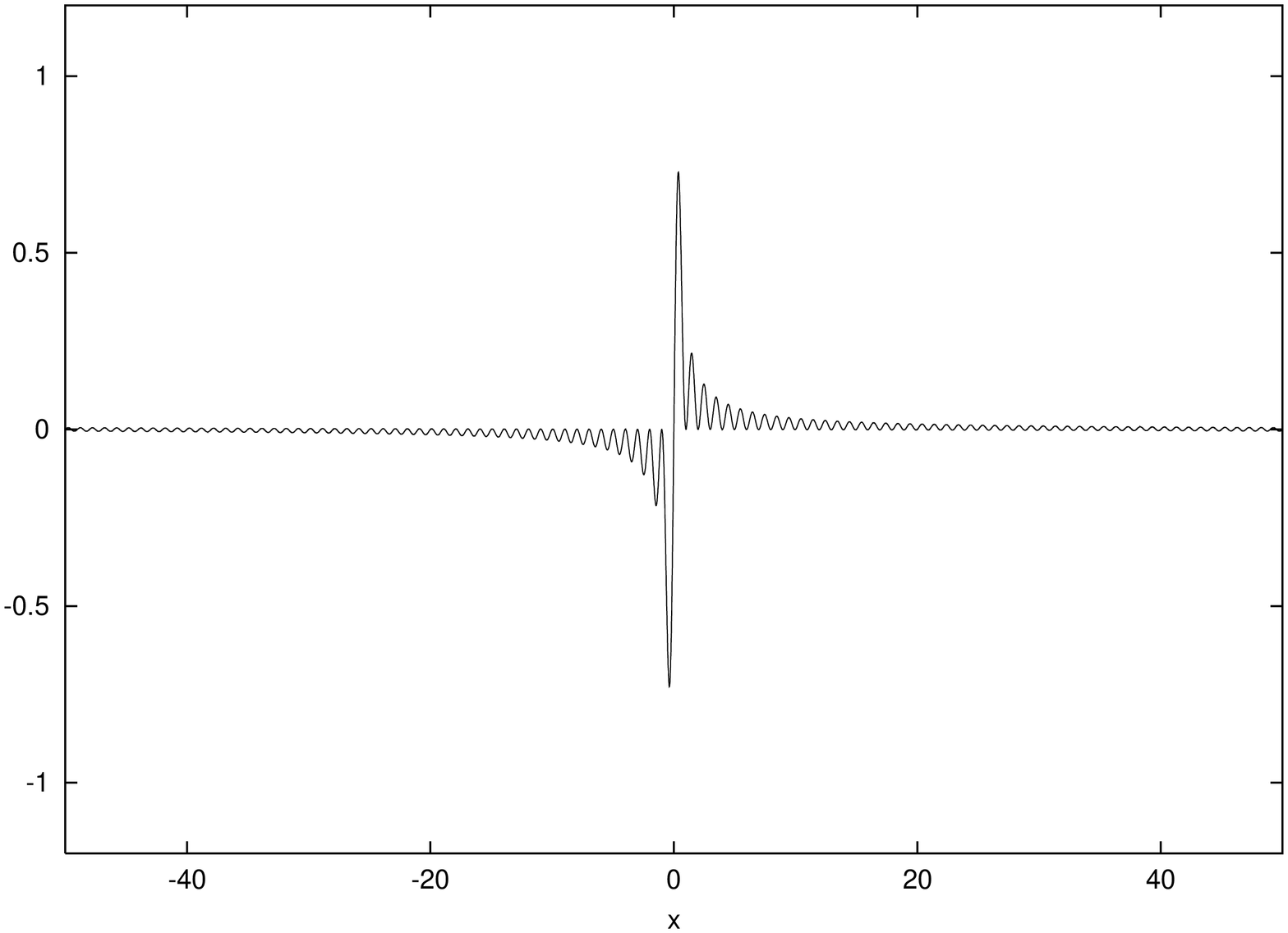,height=12cm,width=6cm,angle=-90} &  &  &  & 
&  &  \\ 
{\large {\bf (b) }} & %
\epsfig{file=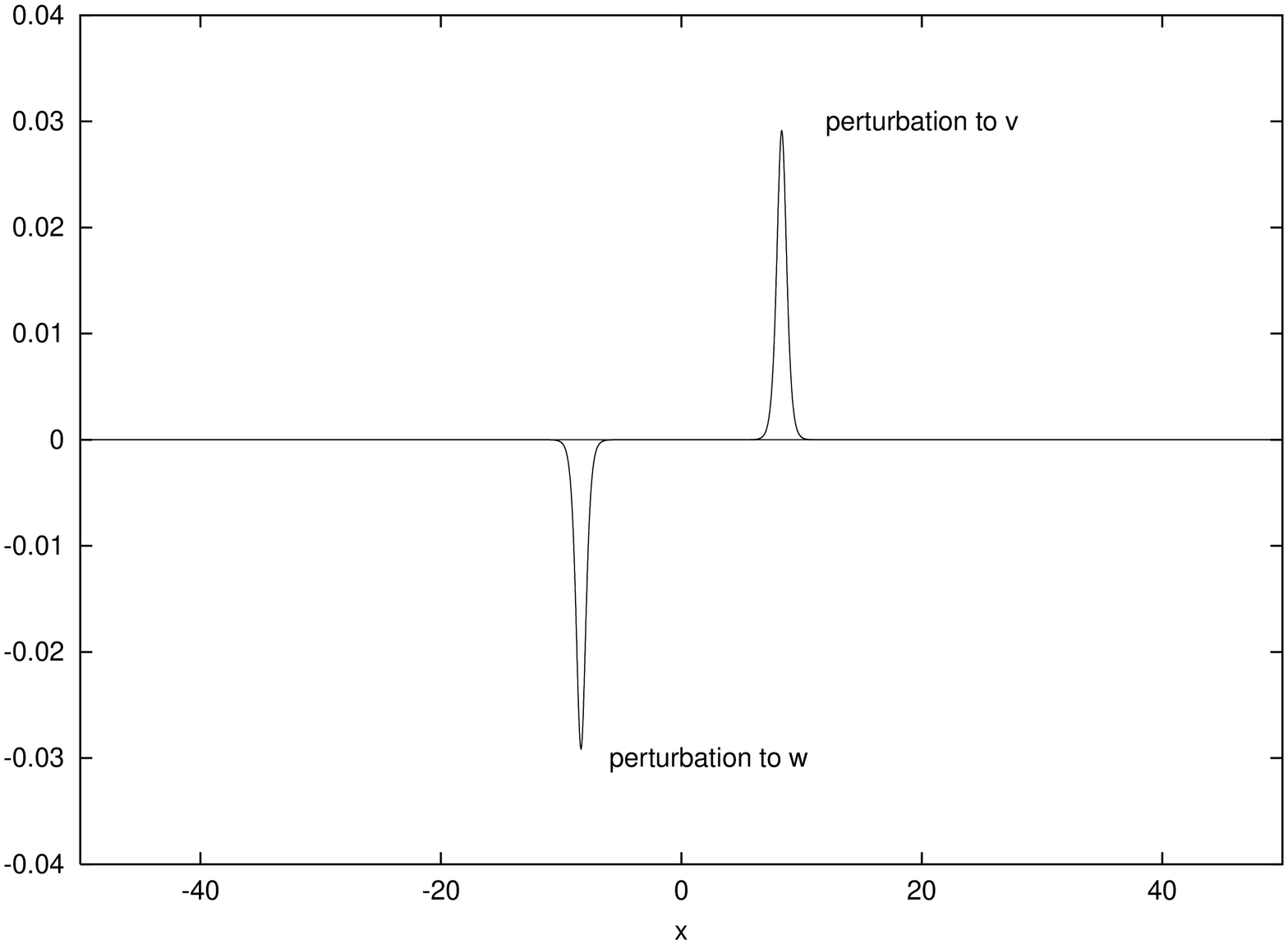,height=12cm,width=6cm,angle=-90} &  &  &  & 
&  & 
\end{tabular}
\caption{(a) A particular example of the initial perturbation added to the
stationary solution of the type B ($v$- and $w$-components of this
perturbation are equal), that gives rise to practically constant residual
perturbation pulses, shown in the panel (b), which are pinned at the points
where the type-B solution makes its jump. In this figure, $a=8.0$, $%
\protect\lambda =2.0$, and the panel (b) pertains to the time moment $t=15$.}
\label{bm05}
\end{figure}

The situation with the type-B solutions is more tricky: perturbations
generally decay, but there remain some residual localized perturbation
pulses, which show a clear trend to be stuck exactly at the spots where the
jumps of the type-B solution are located, see Fig. \ref{bm05}. The fact that
the residual perturbations of the $v$- and $w$-components accumulate at the
point where, respectively, the $v$- and $w$-components of the unperturbed
stationary solution make their jumps, is quite easy to understand, as at
these points both the $v$ and $w$ fields in the unperturbed solution are
nearly equal to zero, hence the nonlinear suppression of the perturbations
does not locally take place, allowing the perturbations to survive.

Further simulations, based on the f{ull} stationary equations (\ref{v}) and (%
\ref{w}), rather than their linearized versions, demonstrate that,
eventually, the residual perturbation pulses seen in Fig. \ref{bm05}(b)
decay, but very slowly. This is natural too, as the above-mentioned
mechanism disabling the nonlinear suppression of the perturbations is only
an approximate one, and it cannot provide for their survival over infinitely
long time.

\section{Conclusion}

In this work, we have introduced a parametrically driven gradient
Ginzburg-Landau model, which is subcritical in the absence of the parametric
drive. In the case when the parametric drive is uniform in space, a kink
solution to this model is stable, on the contrary to the well-known
instability of the same solution in the GL model without the parametric
drive. However, a really nontrivial situation takes places when the
parametric drive is itself subject to the kink-like spatial modulation, so
that it selects real and imaginary constant solutions at $x=\pm \infty $. In
this case, we have found stationary solutions numerically, and also
analytically for a particular case. The solutions look as being of two
different types, viz., a pair of kinks in the real and imaginary components,
or the same with an extra kink inserted into each component. We have showed
that, in fact, both solutions belong to the same continuous one-parameter
family of solutions. Simulations show that the former solution is always
stable, while the latter one admits a special type of small perturbations
that remain virtually constant in time, rather than decaying or growing.
Eventually, these special perturbations decay, but extremely slowly).

We acknowledge valuable discussions with A.A. Nepomnyashchy and W.
Zimmermann, and thank Pablo Funes for his help with the software
implementation. One of the authors (B.A.M.) appreciates hospitality of the
Max Planck Institute for Physics of Complex Systems (Dresden, Germany),
where a part of this work was done. H.G.R. acknowledges support from the
Fischbach Fellowship.

\newpage

\end{document}